\documentclass[aps,prl,twocolumn]{revtex4}

\usepackage{amssymb}
\usepackage{graphicx}
\usepackage{epstopdf}
\usepackage{amsmath}
\usepackage{epsfig}
\usepackage{color}
\usepackage{ulem}

\def\ind#1{{_{\mathrm{#1}}}}

\begin{document}
\title{{\bf Positive-Negative Birefringence in Multiferroic Layered Metasurfaces}}

\author{R. Khomeriki}
\affiliation{Institut f\"ur Physik, Martin-Luther-Universit\"at,
Halle-Wittenberg, D-06099 Halle/Saale, Germany \\ Physics
Department, Tbilisi State University, 3 Chavchavadze, 0128 Tbilisi,
Georgia}
\author{L. Chotorlishvili}
\affiliation{Institut f\"ur Physik, Martin-Luther-Universit\"at, Halle-Wittenberg, D-06099 Halle/Saale, Germany}
\author{I. Tralle}
\affiliation{Faculty of Mathematis and Natural Sciences, University of Rzeszow,   Pigonia  1, 35-310 Rzeszow, Poland}
\author{J. Berakdar}
\affiliation{Institut f\"ur Physik, Martin-Luther-Universit\"at, Halle-Wittenberg, 06099 Halle/Saale, Germany}
\email{Jamal.Berakdar@physik.uni-halle.de}

%

%
%
\begin{abstract}
We uncover and identify the regime for a magnetically and ferroelectrically controllable  negative refraction  of light traversing
 multiferroic, oxide-based metastructure consisting
of alternating nanoscopic  ferroelectric (SrTiO$_2$) and
ferromagnetic (Y$_3$Fe$_2$(FeO$_4$)$_3$, YIG) layers. We perform analytical and numerical simulations based
on discretized, coupled equations for the self-consistent  Maxwell/ferroelectric/ferromagnetic dynamics and obtain
 a biquadratic relation for the refractive index. Various scenarios of ordinary and negative
refraction in different frequency ranges are analyzed and quantified by simple analytical formula that are confirmed by
 full-fledge numerical simulations. Electromagnetic-waves injected  at the edges of the sample are propagated exactly numerically.  We discovered that for  particular  GHz frequencies,  waves with different polarizations are characterized by different signs of the refractive index giving rise to  novel  types of phenomena such as a positive-negative birefringence effect, and magnetically controlled light trapping and accelerations.
  \end{abstract}

  \maketitle
\section{Introduction}
According to Veselago's predictions \cite{ves1,ves2} that were later confirmed
experimentally \cite{pendry1,smith,shelby}, negative
refraction phenomena occur  in metamaterials where both the
electric permittivity $\epsilon$ and the magnetic permeability $\mu$ are negative. The
  Poynting and the wave vectors are then antiparallel  resulting in a
phase decrease during the propagation process. A clear example of this phenomena is
based on metallic heterostructures
\cite{pendry2} which are inherently absorptive at relevant
frequencies \cite{stoc}. To avoid losses, insulating multiferroics may offer a solution \cite{ward,fredkin}, but also  new possibilities
for external control and exploitations of functional materials.
Multiferroics are one-phase or composite, synthesized structures  exhibiting simultaneously  multiple
 orderings such as  ferromagnetic, ferro and/or piezoelectric order
and respond thus to a multitude of conjugate fields.
This class of materials plays a key role for addressing fundamental issues
regarding the interplay  between electronic correlation, symmetry,
magnetism, and polarization. Potential applications are diverse,
ranging from sensorics and magnetoelectric spintronics to
environmentally friendly devices with ultralow energy consumption
\cite{mat1,mat2}.

Here, we demonstrate  how multiferroics properties lead to
 exotic electromagnetic wave propagation features.
 In particular, we demonstrate  the existence of a negative refraction in  ferroelectric (FE)/ferromagnetic(FM)
multilayers.
 The large  (FE and FM) resonance frequency mismatch between  ferroelectric and ferromagnetic
media is usually an obstacle. Indeed, the paradigm ferroelectric BaTiO$_3$ has a resonance
frequency in THz range \cite{ba1,ba2,ba3,ba4}, while the insulating
ferromagnet, rhodium-substituted $\varepsilon$-Rh$_x$Fe$_{2-x}$O$_3$
with largest known coercivity has characteristic frequencies in
200GHz range \cite{fer1}. On the other hand, ferroelectric
SrTiO$_2$ (STO) \cite{nat} does possess overlapping resonances with  the well-investigated
insulating ferromagnet Y$_3$Fe$_2$(FeO$_4$)$_3$ (also called YIG) \cite{lev}. A number of further insulating FE, and FM insulating oxides are
also possible, for concreteness we present and discuss here the results for STO/YIG/STO/$\cdots$ structures. In the pilot numerical simulations
below we choose the FE layer to be 10 $nm$ and the FM layer to be 1 $\mu m$. As we will be working in reduced units, meaning
 the effects are scalable; for materials composites responding at
higher frequencies, smaller structures are appropriate.

\begin{figure}[t]
\centering \includegraphics[scale=.34]{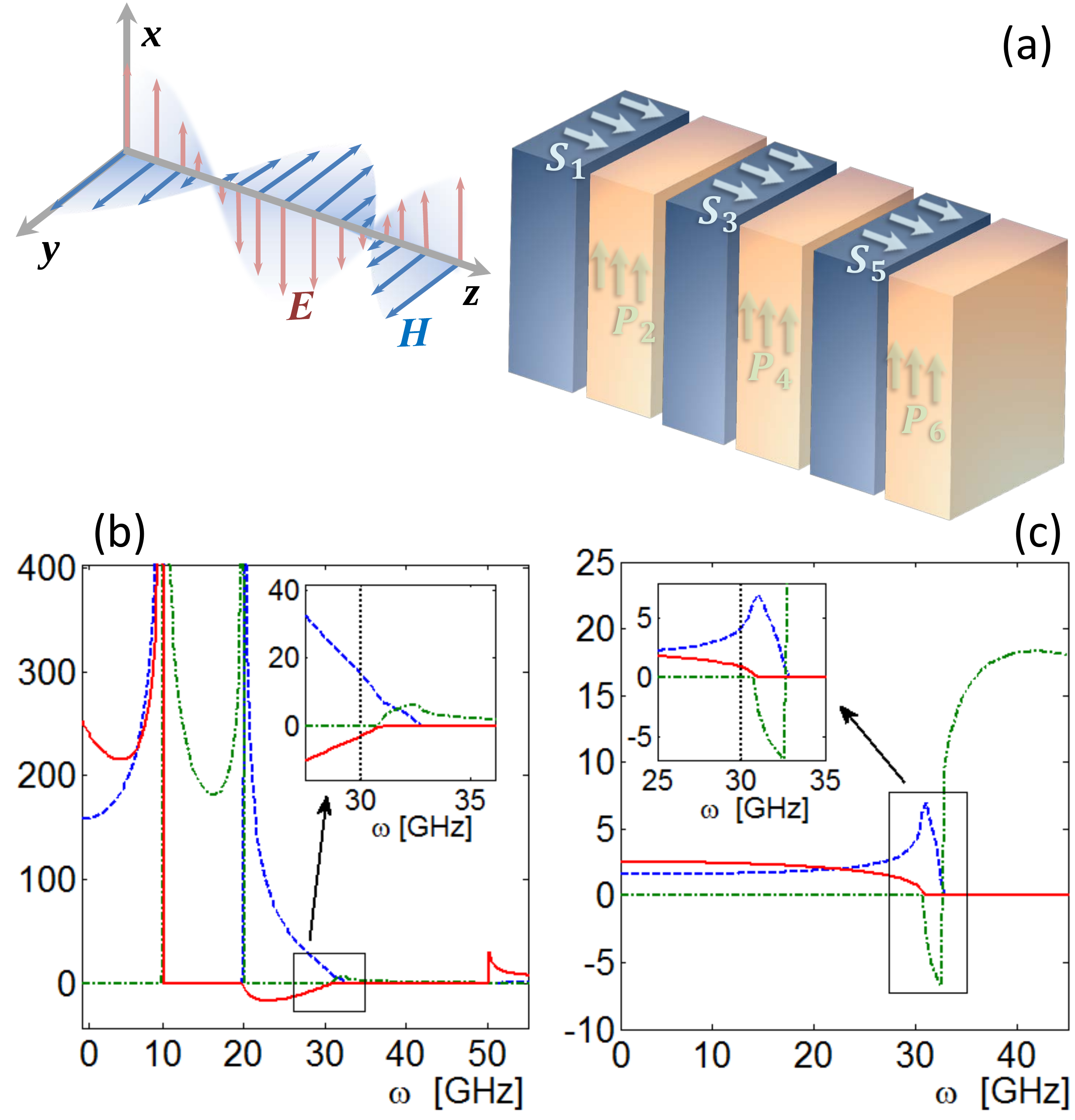} \caption{ (a)
Schematics for the photonic/ferroelectric/ferromagnetic
heterostructure. The electric polarization  ${\mathbf P}_j$ in the
layer $j$ is aligned along $x$ axis. The incident light wave
propagating along  $z$ axis.
 ${\mathbf E}$ and $\mathbf{H}$ denote the electric and magnetic field components.
The magnetization ${\mathbf S}_i$ in the layer $i$  points also along the $z$ axis.
(b) and (c) graphs show real (dashed blue curve) and
imaginary (dashed-dotted green curve) parts of the refraction index\ $n$
versus the mode frequency according to the biquadratic equation \eqref{rel}.
Red solid curve is the $z$ component of the time averaged dimensionless
Poynting vector $\langle W_z\rangle/P_0S_0$ as calculated by means of the
formula \eqref{point}. Graphs (b) and (c) correspond to the negatively
and positively refracting waves, respectively for the same
excitation frequency $30$GHz indicated by vertical dotted lines in
the insets. Insets display enlarged views of  the frequencies in
interest. The FE/FM specific  materials are detailed in the text. }
\label{fig_0}
\end{figure}

In earlier studies the negative refraction effect was observed in
specific systems embracing two different subparts with $\epsilon<0$,
$\mu>0$, and $\epsilon>0$, $\mu<0$ respectively \cite{ward,fredkin}.
We note that in the same medium propagating positively refracting
mode can exist as well \cite{fredkin}.
In the present paper, we study the  FE/FM composite (cf. the schematics setup in Fig.
\ref{fig_0}a) with an eye to
find the frequency domain where both negatively and
positively refracting waves with different polarizations
simultaneously coexist for the same excitation frequency. By this we
predict that in the suggested system  unpolarized electromagnetic wave undergoes both
positive and negative refractions, manifesting a novel
positive-negative birefringence effect.

\section{Model and parameters}

Let both the  thickness of the YIG and the STO layers be equal $a$, for clarity. We denote the positions of STO and
YIG layers by even ($2m$) and odd ($2m-1$) integer numbers
respectively, where $m=1,~2,\dots,~N/2$. For the description of light-induced
FE/FM dynamics and its backaction on the light propagation properties we will utilized
a  discretized  Maxwell materials equation self-consistently  coupled
to  the FE dynamics as described by the  Ginzburg-Landau-Devonshire (GLD) method,
 and a classical Heisenberg model for the magnetization precession.
This low-energy effective treatment is well-justified due to
 the choice of the appropriate frequency  and the (low to moderate) intensity of the incident light wave.
The  discretized FE polarization $P_{2m}$ (initially along $x$ axis) and  FM
magnetization $\vec{S}_{2m-1}$ (initially along $z$ axis) (cf. Fig. \ref{fig_0}a)  are thus described by the
energy functional
\begin{eqnarray}
&&{\cal H} = {\cal H}\ind{P}  + {\cal H}\ind{S}, \qquad {\cal
H}\ind{P}  = \sum\limits_{m = 1}^{N/2} \biggl[\frac{\alpha_0}{2}
\left(\frac{dP_{2m}}{dt}\right)^2 -
\nonumber \\
&& - \frac{\alpha_1}{2}\left(P_{2m}\right)^2 +
\frac{\alpha_2}{4}\left(P_{2m}\right)^4 - P_{2m}E_{2m}^x\biggr],
\label{1} \\
&&{\cal H}\ind{S}  = -\sum\limits_{m
=1}^{N/2}\left[H_0S_z+D\left(S_{2m-1}^z \right)^2
+\vec{H}_{2m-1}\vec{S}_{2m-1}\right], \nonumber
\end{eqnarray}
where $\alpha_0$ stands for the kinetic constant, $\alpha_1$ and
$\alpha_2$ are potential coefficients of the FE part, and $D$ is a uniaxial  anisotropy constant in FM layers. $H_0$ is
a static external magnetic field applied along the $z$ axis
which will prove useful for tuning the functional properties of the setup, e.g.  for switching between the ordinary and the negative refraction regimes (see below).

We assume that the multilayer structure is first driven to saturation by appropriately strong fields. The remnant FE and FM polarizations are
then denoted by   $P_0$ and $S_0$. Let us
introduce dimensionless photonic field
 $\vec{h}\equiv\vec{H}/S_0$, and  $\vec{\cal
E}\equiv\vec{E}/P_0$ and denote small deviations  around the ordering
directions by $p_{2m}\equiv P_0-P_{2m}$, and ${\vec s}_{2m-1}\equiv {\vec
S}_0-{\vec S}_{2m-1}$. The thickness of the layers (along the propagation direction) should be small enough such that no domains are formed
along the $z$ axis
(the general case of large $a$ is captured also with this model by adding pinning sites and appropriate energy contributions to eq.(\ref{1}), but
this is expected to
be subsidiary to the effects discussed here).
Our propagating electromagnetic wave cannot create domains since the wavelength far exceeds
 $a$.
The discretized form of Maxwell's equations for the
electromagnetic field vectors read
\begin{eqnarray}
\frac{1}{c}\frac{d
}{dt}\left(h_{2m+1}^x+4\pi{s}_{2m+1}^x\right)=\frac{1}{2a}\left(
{\cal E}_{2m+2}^y-{\cal E}_{2m}^y\right) \nonumber \\
-\frac{1}{c}\frac{d
}{dt}\left(h_{2m+1}^y+4\pi{s}_{2m+1}^y\right)=\frac{1}{2a}\left(
{\cal E}_{2m+2}^x-{\cal E}_{2m}^x\right) \nonumber \\
-\frac{1}{c}\frac{d }{dt}\left({\cal
E}_{2m}^x+4\pi{p}_{2m}\right)=\frac{1}{2a}\left(
h_{2m+1}^y-h_{2m-1}^y\right) \nonumber \\
\frac{1}{c}\frac{d}{dt}{\cal E}_{2m}^y=
\frac{1}{2a}\left(h_{2m+1}^x-h_{2m-1}^x\right) \label{2} .
\end{eqnarray}
These equations need to be propagated simultaneously with the dynamics governed by eqs.(\ref{1}).
 Nonlinear corrections are irrelevant when the relative values
of the polarization and the magnetization of eigenmodes are much smaller than
unity. Thus, the validity of the linear approximation can be checked directly by monitoring the
relative eigenmodes.  After these simplifications, from \eqref{1} we infer the  linearized, coupled
 photonic-matter evolution equations
\begin{eqnarray}\label{3}
\frac{d^{2} p_{2m}}{d t^{2}}  &=& - \omega_P^2
p_{2m}+\frac{\omega_P^2}{4\pi\alpha}{\cal E}_{2m}^x \\
\frac{\partial {s}_{2m+1}^x}{\partial t} &=&
-\omega_0{s}_{2m+1}^y+\frac{\omega_M}{4\pi}h^y_{2m+1}
\nonumber \\
\frac{\partial {s}_{2m+1}^y}{\partial t} &=&
\omega_0{s}_{2m+1}^x-\frac{\omega_M}{4\pi} h^x_{2m+1}. \nonumber
\end{eqnarray}

The FE resonance frequency  $\omega_P$ of STO is around $\omega_P=10$ GHz. The GDL potential
curvature at equilibrium $\alpha=2\alpha_1$ is related to the electric susceptibility at the zero mode frequency as
$\chi(0)=1/4\pi\alpha$. For the large permittivity observed in Ref. \cite{nat} the potential curvature is of the order of
$\alpha\sim 10^{-4}$. For YIG \cite{lev} the Larmor
frequency is $\omega_0=\gamma H_0+2D\gamma S_0$ (in zero
external field $\omega_0=20$GHz) and $\omega_M=4\pi\gamma
S_0=30$GHz ($\gamma$ is the gyromagnetic ratio for electrons).


For a linearly polarized
electromagnetic waves in the FE/FM multilayers the refractive index
follows from the matter equations \eqref{3}.
The expressions for the  permittivity
$\epsilon=1+\frac{\omega_P^2/\alpha}{\omega_P^2-\omega^2}$ and the
permeability $\mu=1+\frac{\omega_0\omega_M}{\omega_0^2-\omega^2}$
for the linearly polarized wave components ${\cal E}_x$,  $h_y$ indicate
that
\begin{equation}
n^2_1=\epsilon\mu=\left(1+\frac{\omega_P^2/\alpha}{\omega_P^2-\omega^2}\right)\left(
1+\frac{\omega_0\omega_M}{\omega_0^2-\omega^2}\right).
\label{refraction1}
\end{equation}

In spite of the fact that a linearly polarized wave is not an
eigenmode of the FE-FM system, the eigenmode (\ref{refraction1}) has a
certain merit. The asymptotic solution corresponds to the large
susceptibility limit (see below). In order to precisely calculate the
refractive index, one needs to solve  a complete set of coupled Maxwell
\eqref{2} and matter \eqref{3} equations. Thus looking for a
general solution we proceed further and adopt an ansatz presenting
field and matter wave components as follows: ${\cal E}_m^x={\cal
E}_xe^{i(\omega t-kam)}$. Here $a$ is a FE/FM lattice constant
$\omega$ and $k$ are the eigenmode frequency and wavenumber,
respectively. Analyzing the linear algebraic equations (see Supporting Information for
the details) we arrive at the following biquadratic
equation for the refractive index $n\equiv ck/\omega$.
\begin{eqnarray}
\alpha(\omega_0^2-\omega^2)(\omega_P^2-\omega^2)n^4- \label{rel} \\
-(\omega_0^2+\omega_M\omega_0-\omega^2)\left[2\alpha(\omega_P^2-\omega^2)+\omega_P^2\right]n^2+
\nonumber \\
+\left[\alpha(\omega_P^2-\omega^2)+\omega_P^2\right]\left[(\omega_0+\omega_M)^2-\omega^2\right]=0,
\nonumber
\end{eqnarray}
and a set of amplitudes for the field and matter wave components. Then
it is straightforward to calculate the time averaged dimensionless
Poynting vector ${\cal W}=\langle W_z\rangle /P_0S_0$ as
\begin{eqnarray}
{\cal W}=\langle E_xH_y-E_yH_x\rangle/P_0S_0=\left[{\cal
E}_xh_y^*-{\cal E}_yh_x^*\right]+c.c.  \nonumber \\
=n\left\{\frac{(\omega^2-\omega_P^2)\left[(\omega^2-\omega_0^2)(n^2-1)+\omega_0\omega_M\right]^2}
{\left(\omega^2-\omega_P^2-\omega_P^2/\alpha\right)\omega^2\omega_M^2}+1\right\}.
\label{point}
\end{eqnarray}

{It is straightforward to derive from \eqref{rel} the two obvious  limiting cases of
pure ferromagnet or ferroelectrics. For zero
magnetization, we set $\omega_M=0$ and from \eqref{rel}
obtain  two linearly polarized eigenmodes with the refractive indexes
$n_1=1$ and
$n_2=\sqrt{1+\frac{\omega_P^2/\alpha}{\omega_P^2-\omega^2}}$
corresponding to the polarizations along $y$ and $x$ axis, respectively.
In the case of a vanishing polarization in the system we  set
$\alpha\rightarrow\infty$ and find two circularly polarized
eigenmodes characterized by different refractive indexes
$n_1=\sqrt{1+\frac{\omega_M}{\omega_0-\omega}}$ and
$n_2=\sqrt{1+\frac{\omega_M}{\omega_0+\omega}}$.

\begin{figure}[t]
\centering \includegraphics[scale=.6]{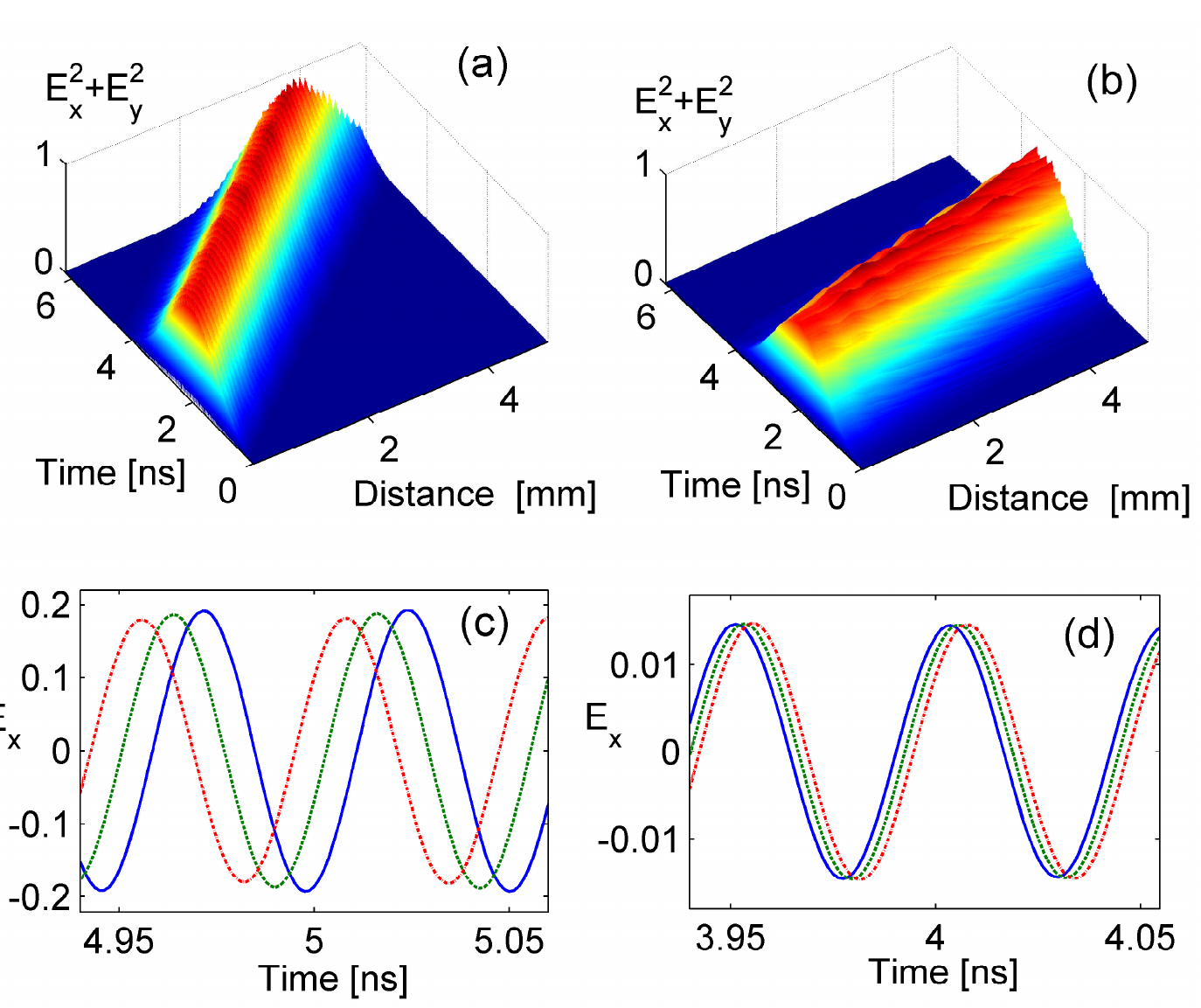} \caption{Graphs (a)
and (c) represent negative refraction injecting the signal with a
polarization ${\cal E}_{x,y}=0.19+1.81i$. while Graphs (b) and (d)
display positive refraction of the wave with polarization ${\cal
E}_{x,y}=0.015+1.985i$ corresponding to a positive refraction. In
graphs (c) and (d) blue (solid), green (dashed-dotted) and red
(dashed) curves correspond to time oscillations of the FE/FM layers
with increasing site numbers.} \label{fig_1}
\end{figure}
Obviously all of these modes in both limiting cases are
characterized by ordinary refraction properties for low excitation
frequencies $\omega$, while for large $\omega$ one of refractive
indexes becomes purely imaginary corresponding to the non-propagating
regime, while other remains real and corresponds to an ordinary
refraction. It should be  emphasized that in the above
mentioned cases with the linear polarized wave at the input
always  obtain linearly polarized wave at the output: In purely
ferromagnetic case we find a Faraday rotation and for pure
ferroelectrics just a phase shift (that however can be retrieved by interference measurements).\\
 The situation  changes drastically in the
presence of both ferroelectric and ferromagnetic layers.
An injected  linear polarized electromagnetic wave  emerges after traversing the FE/FM heterostructure with
 an elliptical polarization, as detailed below.}

{We can further simplify the analytic solutions \eqref{rel} and
\eqref{point} in the general case (both ferroelectric and ferromagnetic
layers are present in the system) assuming $\alpha\rightarrow 0$
which is just the case of a large susceptibility  in
FE \cite{nat}. Then one infers two roots of
\eqref{rel}. The first one  matches exactly the asymptotic result
\eqref{refraction1}, and the second one reads
\begin{equation}
n_2^2=\frac{(\omega_0+\omega_M)^2-\omega^2}{\omega_0(\omega_0+\omega_M)-\omega^2}.
\label{refraction2}
\end{equation}

Now it is evident  that in the same limit of $\alpha\rightarrow 0$, for
the first mode \eqref{refraction1} the Poynting vector has the
values ${\cal W}_1\sim -n$, meaning that a negative refraction takes place.
 For the second mode \eqref{refraction2} the Poynting vector is
${\cal W}_2\sim n$ corresponding to a positive refraction case. To
identify the positive-negative birefringence regime
both propagating modes should be present, i.e. $n_1^2>0$ and $n_2^2>0$. From these
relations we deduce the restrictions on the mode frequency
\begin{equation}
\texttt{Max}\{\omega_P,\omega_0\}<\omega<\sqrt{\omega_0(\omega_0+\omega_M)}.
\label{cond}
\end{equation}
These relations set a reliable  range in which  positive-negative
birefringence effect  to be found.

\section{Results}

The exact solutions of \eqref{rel} are illustrated  by
graphs (b) and (c) in Fig. \ref{fig_0}. The frequency
dependencies of the two roots with positive real parts $n^\prime>0$ of
the refractive index $n=n^\prime+n^{\prime\prime}$ are displayed as to pinpoint the
wavevector direction and compare it with the sign of the Poynting vector as
calculated according to \eqref{point}. Apparently in Fig.
\ref{fig_0} (b), the sign of the Poynting vector is negative in the
frequency range close to $\omega=30$GHz, i.e. the Poynting vector is
antiparallel to the wave vector direction and therefore we are in the
negative refraction regime, while in graph (c) another root with a
positive $n^\prime$ is presented. The Poynting vector in this case is
positive, that means we have a second mode with an ordinary
(positive) refraction for the same wave frequency $\omega=30$GHz.
The clear evidence  of the coexistence of a positive and negative refractions
for an unpolarized electromagnetic wave is fully compatible with the
approximate conditions \eqref{cond}.

To substantiate the analytical estimation we performed full numerical
experiment considering two wave modes with different polarizations being
injected into the FE/FM composite metastructure. An oscillating electric field
with the characteristic frequency $\omega=30$GHz operating  on the left
edge of the dipolar/spin chain is due to the  action of the light source on
the sample. The wave propagation proceeds  self-consistently as governed by the
set of equations \eqref{2} and \eqref{3}. The results shown in Fig.
\ref{fig_1} confirm evidently  the existence of the birefringence
effect.

In numerical simulations we act on the left end of the system with an
electric field having the polarization ${\cal E}_{x,y}=0.19+1.81i$ and
corresponding to the negative refraction regime. Obviously a Gaussian
pulse propagates with a positive group velocity see Fig. \ref{fig_1}
(a), while according to Fig. \ref{fig_1} (c) the phase velocity is
negative (blue, green and red curves correspond to time oscillations
of subsequent sites with increasing site number). Fig. \ref{fig_1}
(b) displays  the wave propagation process (again with positive group
velocity) with the polarization ${\cal E}_{x,y}=0.015+1.985i$. However,
 the corresponding phase velocity is now positive and ordinary (positive)
refraction scenario holds, see Fig. \ref{fig_1} (d).

In the  above considerations the damping effects were
neglected which allows obtaining  analytical
expressions. In practice,   ferromagnetic layers can be engineered such that the  damping \cite{damp}
is very small (in $\mu$s range) and hence can neglected on our relevant
$n$s time scale. In ferroelectrics damping is much stronger and its effect should be considered.
FE damping impacts the first
mode only \eqref{refraction1}. In the limit $\alpha\rightarrow 0$
(i.e. for large susceptibility) the modified expression of the
refraction index of the first mode reads
\begin{equation}
n^2_1=\left(1+\frac{\omega_P^2/\alpha}{\omega_P^2-\omega^2+i\omega\Gamma}\right)\left(
1+\frac{\omega_0\omega_M}{\omega_0^2-\omega^2}\right),
\label{refraction11}
\end{equation}
(here $\Gamma$ is the damping parameter) while the second  mode
\eqref{refraction2} is left unchanged. The experimentally observed peaks in the susceptibility
correspond to the frequency $\omega_p=10$GHz see Ref \cite{nat}.
In our case $\omega=30$GHz, $\Gamma\ll\omega_P\ll\omega$. Hence, we conclude that
also FE damping has no significant effect on the first mode  \eqref{refraction11} clarifying so the role of losses for the predicted phenomena.
Until now we considered FM and FE layers of equal thickness.
Absorption effects are rather sensitive to the thickness of FE
layers and much less sensitive to the thickness of FM layers. A recipe
for minimizing losses  is  thus straightforward by fabricating thinner
FE layers. In the numerical simulations the
thickness enters through  the values of $\alpha$
in  \eqref{3}. For example, taking $\alpha=10^{-2}$ instead of
$\alpha=10^{-4}$ in \eqref{3}, one can make the average polarization
100 times smaller (which affects the electromagnetic waves). If we
take the FE layer 100 times thinner than the FM layer, then the wave spends much
less time in FE layer and the absorption effects are reduced drastically.
On the other hand, tuning $\alpha$ leaves the main qualitative
characteristics of the considered effect unchanged.

The frequency range for which the negative refraction takes place
depends on the external magnetic field  (see Fig. \ref{fig_2}).
The external magnetic field induces a shift in the Larmor frequency
$\omega_0=\omega_a+\gamma H_0$, while the anisotropy frequency is fixed to
$\omega_a=20$GHz. By means of an external magnetic field the
frequency $\omega_0$ is tunable within an interval $10$GHz to $50$GHz. The
results for the negative refraction are shown in Fig. \ref{fig_2}
(left graph). These results confirm that the frequency range for
which a negative refraction takes place can be controlled by an external
magnetic field. In the experiment one may switch so negative
refraction media to ordinary media and vice versa simply by  turning
the magnetic field  off and on.

\begin{figure}[t]
\centering \includegraphics[scale=.6]{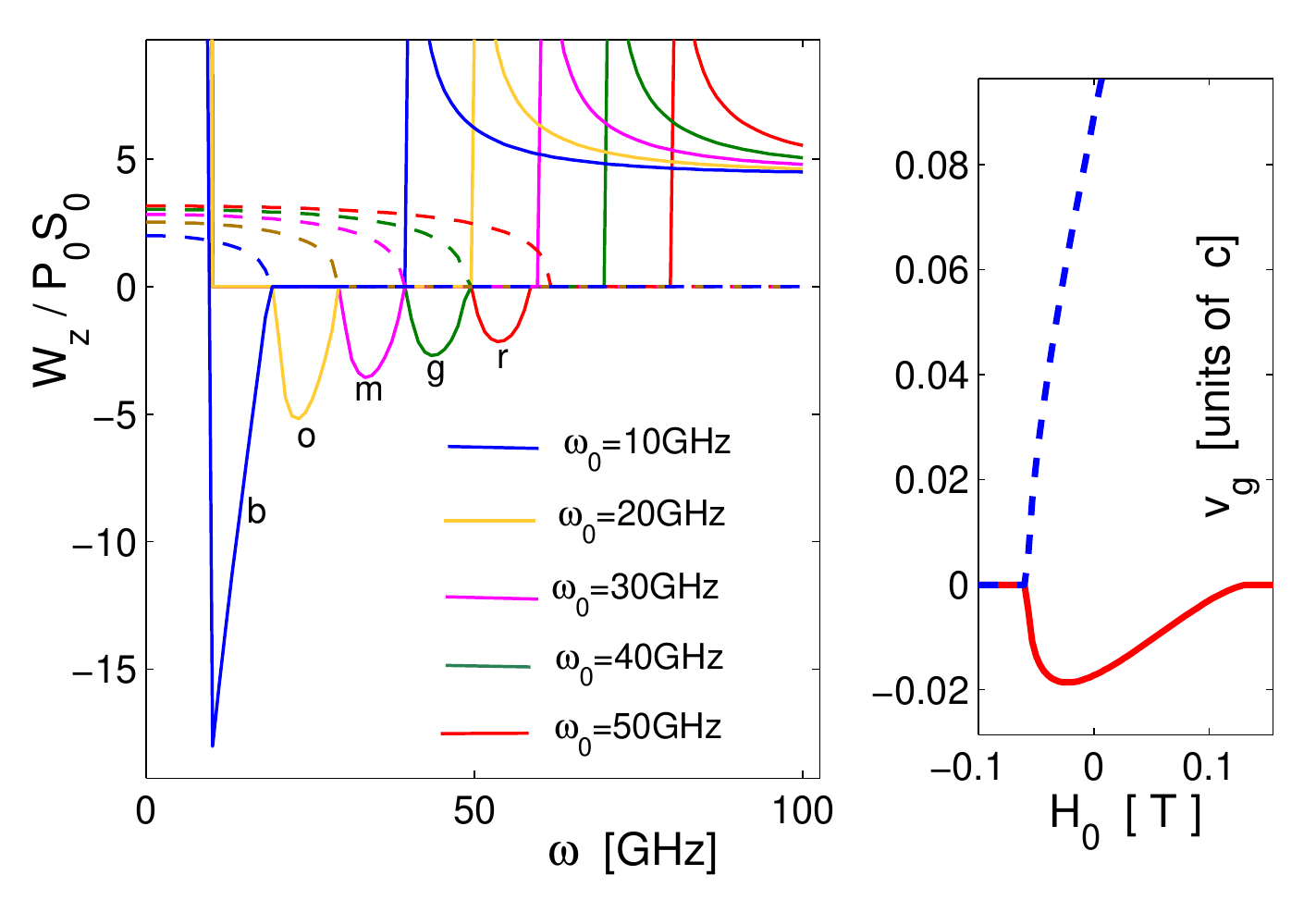} \caption{Left
graph: The Poynting vector magnitude  versus the mode frequency for
different external magnetic fields: $\omega_0=10,~20,~30,~40,~50$
GHz correspond to blue (b), orange (o), magenta (m), green (g), and
red (r) lines. Solid curves indicate the modes with a negative
refraction regime and dashed lines describe the positively
refractive ones. The frequency ranges with a vanishing  Poynting
vector  correspond to non-propagating evanescent modes. {Right
graph: The mode group velocity dependence on the static magnetic
field for a fixed excitation frequency $\omega=24$GHz. Blue (dashed)
and red (solid) curves describe the group velocities (in units of
the speed of light) of positively and negatively refracting modes,
respectively.}} \label{fig_2}
\end{figure}

In the right graph of Fig. \ref{fig_2} for negatively and positively
refracting waves we plot the dependence of the group velocity on a static magnetic field.
It is evident that by a ramped static magnetic field we can achieve acceleration or even trapping of the electromagnetic
waves.
Finally we note that effects related to photonic-magneto-elastic and/or piezoelectric couplings are straightforwardly incorporated in the above
formulism by  including the respective energy term in eqs.(\ref{1}), along the lines as done in Refs.\cite{jia,jia2}.
Furthermore, to access a  higher frequency regime (cf. eq.\ref{refraction2})) it would be advantages to utilize FE/aniferromagnetic/FE/.... layer structures.
For example, recently an antiferromagnetic resonance frequency of  22 THz  were observed for KNiF$_3$  \cite{natcomm}.

\section{Summary}

Summarizing,  in the present work we   illustrated theoretically an insulating  multiferroic
 metamaterial featuring  simultaneous
 positive and negative refraction (positive-negative birefringes) and provided
 concrete predictions for  a  realization on the basis of
   $SrTiO_2/YIG$ multilayers. In addition to  full-fledge numerical simulations for the coupled
   Maxwell/ferroelectric/ferromagnetic dynamics, we were able to derive credible analytical
   solutions and concrete frequency regimes in which the predicted effects are to be expected.
   The theory and predictions are of a general nature and are applicable to a wide range of
   material classes. The findings point to new exciting applications of  insulating  nanostructured
   oxides in photonics.

\begin{acknowledgements}
This research is funded by the German Science foundation under  SFB 762 "Functionality of
Oxide Interfaces". We thank J. Schilling for comments on the
experimental aspects.
R. Kh. acknowledges financial support from Georgian SRNSF (grant No FR/25/6-100/14)
and travel grants from Georgian SRNSF and CNR, Italy (grant No 04/24) and CNRS, France (grant No 04/01).
\end{acknowledgements}

\section{Appendix}

To obtain  wave solutions of the set Eqs.(2,3) in the main text we
express the field and the polarization/magnetization components as
follows:
\begin{eqnarray}
&{\cal E}^x_m={\cal E}_xe^{i(\omega t-kam)}+c.c, \quad {\cal
E}^y_m={\cal E}_ye^{i(\omega t-kam)}+c.c, \nonumber \\
&h_m^x=h_xe^{i(\omega
t-kam)}+c.c, \quad h^y_m=h_ye^{i(\omega t-kam)}+c.c, \nonumber \\
&s_m^x={\cal S}_xe^{i(\omega t-kam)}+c.c, \quad s_m^y={\cal
S}_ye^{i(\omega t-kam)}+c.c, \nonumber \\
&p_m={\cal P}e^{i(\omega t-kam)}+c.c. \label{111}
\end{eqnarray}
Substituting this into Eqs.(2,3)  of the main text and considering
the large wavelength limit $k\rightarrow 0$ we find
\begin{eqnarray}
&i\omega\left(h_x+4\pi{\cal S}_x\right)=-ik{\cal E}_y, \quad
-i\omega\left(h_y+4\pi{\cal S}_y\right)=-ik{\cal E}_x
\nonumber \\
&-i\omega\left({\cal E}_x+4\pi{\cal P}\right)=-ikh_y, \quad
i\omega{\cal
E}_y=-ikh_x \nonumber \\
&i\omega{\cal S}_x+\omega_0{\cal S}_y-\frac{\omega_M}{4\pi}h_y=0
\quad
i\omega{\cal S}_y-\omega_0{\cal S}_x+\frac{\omega_M}{4\pi}h_x=0 \nonumber \\
&\alpha\left(\omega^2-\omega_P^2\right){\cal
P}+\frac{\omega_P^2}{4\pi}{\cal E}_x=0. \label{112}
\end{eqnarray}

After some algebra one can reduce this equation to the three coupled
equations
\begin{eqnarray}
&i\omega(n^2-1)h_x+\left[\omega_0(n^2-1)-\omega_M\right]
h_y-\omega_0n{\cal
P}=0, \nonumber\\
&-\left[\omega_0(n^2-1)-\omega_M\right]
h_x+i\omega(n^2-1)h_y-i\omega n{\cal P}=0, \nonumber\\
&\omega_P^2nh_y+\left[\alpha(\omega^2-\omega_P^2)
-\omega_P^2\right]{\cal P}=0, \label{113}
\end{eqnarray}
where $n\equiv ck/\omega$ is a refractive index. Thus finally we
arrive at the matrix
\begin{eqnarray}
\left(
  \begin{array}{ccc}
  ~~~ i\omega(n^2-1) ~~~~~~~~ \omega_0(n^2-1)-\omega_M ~~~~~~~ \omega_0n ~~~~~ \\
    \omega_M-\omega_0(n^2-1) ~~~~~~~~ i\omega(n^2-1) ~~~~~~~~~ -i\omega n~~~~~~\\
  ~~~~~~~~~~~  0 ~~~~~~~~~~~~~~~ \omega_P^2n ~~~~~~~~~~ \alpha(\omega^2-\omega_P^2)
-\omega_P^2 \\
  \end{array}
\right) \nonumber
\end{eqnarray}
the determinant of which should be equal to zero which leads to the
relation
\begin{eqnarray}
\left[(\omega^2-\omega_P^2)\alpha-\omega_P^2\right]
\left\{\left[\omega_0(n^2-1)-\omega_M\right]^2-\omega^2(n^2-1)^2\right\}
\nonumber \\
+\omega_P^2n^2\left[(\omega_0^2-\omega^2)(n^2-1)-\omega_M\omega_0\right]=0
~~~~~~~~~~~~~~~~~~~~\nonumber
\end{eqnarray}
This could be straightforwardly simplified  to a biquadratic
equation for the refractive index $n$
\begin{eqnarray}
\alpha(\omega_0^2-\omega^2)(\omega_P^2-\omega^2)n^4- \nonumber \\
-(\omega_0^2+\omega_M\omega_0-\omega^2)\left[2\alpha(\omega_P^2-\omega^2)+\omega_P^2\right]n^2+
\nonumber \\
+\left[\alpha(\omega_P^2-\omega^2)+\omega_P^2\right]\left[(\omega_0+\omega_M)^2-\omega^2\right]=0
\nonumber
\end{eqnarray}
which is exactly the same relation (5)  as in the main text.

The above matrix together with the relations \eqref{112} of this
supplementary materials gives the eigenmodes of the system and
defines the time averaged Poynting vector value as
\begin{equation}
\langle W^z\rangle=\langle E_xH_y-E_yH_x\rangle=P_0S_0\left[{\cal
E}_x(h_y)^*-{\cal E}_y(h_x)^*\right]+c.c. \nonumber
\end{equation}
Defining now the dimensionless version of the Poynting vector as
${\cal W}=\langle W_z\rangle /P_0S_0$ we obtain eq.(6) of the main
text
\begin{eqnarray}
{\cal W}
=n\left\{\frac{(\omega^2-\omega_P^2)\left[(\omega^2-\omega_0^2)(n^2-1)+\omega_0\omega_M\right]^2}
{\left(\omega^2-\omega_P^2-\omega_P^2/\alpha\right)\omega^2\omega_M^2}+1\right\}
\nonumber
\end{eqnarray}
which in the limit $\alpha\rightarrow 0$ gives the following
expressions for the negatively refracting $n=n_1$ and positively
refracting $n=n_2$ modes
\begin{eqnarray}
{\cal
W}_1=-n\frac{(\omega_0^2+\omega_0\omega_M-\omega^2)^2\omega_P^2}
{\alpha(\omega^2-\omega_P^2)\omega^2\omega_M^2}, \nonumber \\
{\cal
W}_2=n\left\{1-\frac{\alpha(\omega^2-\omega_P^2)\omega^2\omega_M^2}
{(\omega_0^2+\omega_0\omega_M-\omega^2)^2\omega_P^2}\right\}
\nonumber
\end{eqnarray}
and as it could be easily seen ${\cal W}_1\sim -n$ for
$\omega>\omega_P$ and ${\cal W}_2\sim n$ for $\alpha\rightarrow 0$.

\end{document}